\documentclass[11pt,a4paper]{article}
\usepackage{jheppub}

\usepackage{color}
\usepackage{amsfonts}
\usepackage{slashed}

\usepackage{enumerate}

\def\bea{\begin{eqnarray}}
\def\eea{\end{eqnarray}}
\def\nn{\nonumber}

\def\lmatrix{\left(\begin{array}}
\def\rmatrix{\end{array}\right)}

\def\msbar{\overline{\rm MS\kern-0.5pt}\kern0.5pt}

\long\def\hidestart#1\hideend{}
\def\ol{\overline}

\title{Electroweak interactions and dark baryons in the sextet BSM model with a composite Higgs particle}

\author[abc]{Zoltan Fodor,}
\author[d]{Kieran Holland,}
\author[e]{Julius Kuti,}
\author[cf]{Santanu Mondal,}
\author[cf]{Daniel Nogradi}
\author[a]{and Chik Him Wong}

\affiliation[a]{University of Wuppertal, Department of Physics, Wuppertal D-42097, Germany}
\affiliation[b]{J\"ulich Supercomputing Center, Forschungszentrum J\"ulich, J\"ulich D-52425, Germany}
\affiliation[c]{E\"otv\"os University, Institute for Theoretical Physics, Budapest 1117, Hungary}
\affiliation[d]{University of the Pacific, 3601 Pacific Ave, Stockton CA 95211, USA}
\affiliation[e]{University of California, San Diego, 9500 Gilman Drive, La Jolla, CA 92093, USA}
\affiliation[f]{MTA-ELTE Lendulet Lattice Gauge Theory Research Group, 1117 Budapest, Hungary}

\emailAdd{fodor@bodri.elte.hu}
\emailAdd{kholland@pacific.edu}
\emailAdd{jkuti@ucsd.edu}
\emailAdd{santanu@bodri.elte.hu}
\emailAdd{nogradi@bodri.elte.hu}
\emailAdd{cwong@uni-wuppertal.de}

\abstract{The Electroweak interactions of a strongly coupled gauge theory are discussed
with outlook beyond the Standard Model (BSM) under global and gauge anomaly constraints. 
The theory is
built on a minimal massless fermion doublet of the SU(2) BSM flavor group (bsm-flavor) 
with a confining 
gauge force at the TeV scale in the two-index symmetric (sextet) representation 
of the BSM SU(3) color gauge group (bsm-color).
The intriguing possibility of near-conformal sextet gauge dynamics could lead to the
minimal realization of the composite Higgs mechanism with a light $0^{++}$ 
scalar, far separated from strongly coupled resonances of the confining gauge force
in the 2-3 TeV range, distinct from Higgsless Technicolor.
In previous publications we have presented results for the meson 
spectrum of the theory, including the light composite scalar, perhaps the emergent Higgs impostor. 	
Here we discuss the critically important role of the baryon spectrum in the sextet model investigating its
compatibility with what we know about thermal evolution of the early Universe 
including its galactic and terrestrial relics.
For an important application, we report the first numerical results on the baryon spectrum 
of this theory from non-perturbative lattice simulations with baryon correlators in the
staggered fermion implementation of the strongly coupled gauge sector. 
The quantum numbers of composite baryons and their spectroscopy from lattice simulations are
required inputs for exploring dark matter contributions of the sextet BSM model, as outlined for future work.
}	


\begin{document}

\maketitle

\section{Introduction}
\label{sec:intro}

An important strongly coupled near-conformal gauge theory built on the minimally required
SU(2) bsm-flavor doublet of two massless fermions, with a
confining gauge force on the TeV scale in the sextet
representation of the new SU(3) bsm-color is frequently discussed as an intriguing possibility 
for the minimal realization of the composite Higgs mechanism.
Early discussions of the model as a BSM candidate were initiated in
systematic explorations of higher fermion representations of color gauge groups~\cite{Dietrich:2005jn,Sannino:2004qp,Hong:2004td}
for extensions of the original Higgsless Technicolor paradigm~\cite{Susskind:1978ms,Weinberg:1979bn}.
In fact, the first appearance of the particular two-index symmetric SU(3) fermion representation 
can be traced even further back to Quantum Chromodynamics (QCD) where a doublet of sextet quarks
was proposed as a mechanism for Electroweak symmetry breaking (EWSB)
without an elementary Higgs field~\cite{Marciano:1980zf}. This idea had to be
replaced by a new
gauge force at the TeV scale, orders of magnitude stronger than in QCD, to facilitate the dynamics 
of EWSB just below the lower edge of the conformal window in 
the new BSM paradigm~\cite{Dietrich:2005jn,Sannino:2004qp,Hong:2004td}.
It should be noted that throughout its early history the important near-conformal behavior 
of the model was a hypothesis only and definitive results had to wait for recent
non-perturbative investigations with lattice gauge theory methods
as used in our work.

Near-conformal BSM theories raise
the possibility of a light composite scalar, perhaps a Higgs impostor, to emerge from new strong dynamics,
far separated from the associated composite resonance spectrum in the few TeV mass range
with interesting and testable predictions for the Large Hadron Collider (LHC). 
This scenario is very different from what was expected from QCD when scaled up to the Electroweak scale,
as illustrated by the failure of the Higgsless Technicolor paradigm.
Given the discovery of the 125 GeV Higgs particle at the LHC, any realistic BSM theory must contain a Higgs-like state, 
perhaps with some hidden composite structure.

Based on our {\it ab initio} non-perturbative  lattice calculations
we find
accumulating evidence for near-conformal behavior in the sextet theory with
the emergent low mass  $0^{++}$ scalar state far separated from 
the composite resonance spectrum of bosonic excitations in the 2-3 TeV energy range~\cite{Fodor:2012ty,Fodor:2012ni,Fodor:2014pqa,Fodor:2015vwa}. 
%
The identification of the light scalar state is numerically challenging since it
requires the evaluation of disconnected fermion loop contributions to correlators with vacuum quantum numbers
in the range of light fermion masses we explore.
The evidence to date is very promising that the $0^{++}$ scalar is light 
in the chiral limit and that the model at this stage remains an important BSM candidate.

In Section 2 critically important features of the strongly coupled sextet gauge sector 
of the light scalar are briefly reviewed.
In Section 3 we discuss the Electroweak interactions of sextet fermions and their Electroweak multiplet structure.
The outlook beyond the Standard Model under global and gauge anomaly constraints is presented in Section 4 
including the choice for new leptons under the requirement of integer electric charges for dark baryons.
In Section 5 dark baryons from the sextet Electroweak multiplet structure are constructed 
with discussion of model constraints based on galactic and terrestrial relic densities from the early Universe.
Section 6 describes the construction of lattice baryon operators using staggered lattice fermions 
in the sextet color representation. Our first non-perturbative lattice results on sextet baryon spectroscopy
are presented in Section 7.  We conclude in Section 8 with a brief summary and outlook.

\section{Two critical features of the sextet strong force}
The foundation of the theory is based on chiral symmetry breaking (${\rm \chi SB}$) from the sextet gauge force
in the massless fermion limit with three Goldstone bosons 
for the minimal realization of the Higgs mechanism.
In our work from lattice simulations, the ${\rm \chi SB}$ pattern 
${\rm SU(2)_L\!\otimes\! SU(2)_R\rightarrow SU(2)_V}$  in the  bsm-flavor group
is consistent with the absence of any 
evidence for a conformal infrared fixed point (IRFP) at scales
reached so far with the scale-dependent strong gauge coupling~\cite{Fodor:2015zna}. 
These two fundamental features are intrinsically interdependent.
The existence of an IRFP would make the sextet theory conformal with unbroken chiral symmetry
implying the disappearance of the finite temperature ${\rm \chi SB}$ transition of massless fermions
in the continuum limit~\cite{Kogut:2015zta}.
The phenomenological relevance of the sextet BSM model would be questioned in this case, perhaps with
speculative changes from previously unexplored new features, like the role of four-fermion operators 
in strongly coupled gauge dynamics.
The current understanding of the sextet theory shows no evidence for the paradigm shift 
from an IRFP as briefly argued below.

\subsection{The scale-dependent renormalized coupling and its $\beta$-function}
There have been several lattice studies of the renormalized sextet gauge coupling and its $\beta$-function 
using different schemes. The first studies using the Schr\"{o}dinger functional method with tuned
massless Wilson fermions 
were not decisive to rule in or rule out an IRFP in the explored range 
of the renormalized coupling~\cite{DeGrand:2010na, Shamir:2008pb}. 
Our recent study using the gradient 
flow scheme with exactly massless staggered fermions and with full investigation 
of systematic effects in taking the continuum limit shows no evidence that the $\beta$-function 
has an infrared fixed point in the gauge coupling range ${\rm 0 < g^2 < 6.5}$ investigated~\cite{Fodor:2015zna}.  
This finding is consistent 
with studies of the mass-deformed Goldstone spectrum and the spectrum of the Dirac operator which exhibits
the Banks-Casher condensate in the chiral limit~\cite{Banks:1979yr,Fodor:2015vwa}.
Our ongoing investigations include detailed studies of the mass-deformed Goldstone spectrum
and the chiral condensate via the GMOR relation in chiral perturbation theory. 
Predictions for ${\rm \chi SB}$ from Random Matrix Theory are being tested from the lowest eigenvalues of the Dirac spectrum for high precision results.

Preliminary results from concurrent studies of the scale-dependent renormalized coupling,
using the gradient flow method with
Wilson fermions in finite volumes, reported a conformal IRFP in the sextet $\beta$-function 
within the ${\rm g^2 \sim 5.5-6.5}$ range where the 3-loop and 4-loop perturbative 
$\beta$-functions  develop zeros in the ${\rm \overline{MS}}$ scheme~\cite{Anna_g2}.
These preliminary results have been revised and no IRFP 
is reported anymore in the sextet $\beta$-function~\cite{Hasenfratz:2015ssa}, 
similar to our findings~\cite{Fodor:2015zna}.

It should be noted that the precise determination of the very small $\beta$-function 
presents challenges even for the best gradient flow based methods which 
were deployed by both groups under discussion.
Nevertheless, the outcome of these difficult scale-dependent gauge coupling studies
remains consistent with our expectation that the theory is very close 
to the lower edge of the conformal window with ${\rm \chi SB}$ but without an IRFP.

\subsection{Finite temperature chiral transition}
Based on the existence of  ${\rm \chi SB}$ at zero temperature,
it would be expected that chiral symmetry is restored in the sextet theory in a finite-temperature 
chiral transition. This has been the focus 
of recent work with evidence presented for a chiral transition 
at finite lattice cutoff~\cite{Kogut:2011ty,Kogut:2010cz,Kogut:2015zta}. 
Tracking the cutoff-dependent temperature of the chiral transition as the gauge coupling is varied,
the authors conclude that the chiral transition should disappear in the continuum limit 
and they report new-found indications for a conformal IRFP in the continuum model~\cite{Kogut:2015zta}. 
This finding is based on the scale-dependent variation of the $\beta$-function with the bare and renormalized gauge
couplings, significantly slower than the expected 2-loop perturbative behavior 
in the renormalized weak coupling range ${\rm g^2_R \sim 1-3}$ but without the removal of cutoff effects
in the $\beta$-function and without control of other systematics which can qualitatively affect the conclusions of~\cite{Kogut:2015zta}.
In particular, the simple Wilson gauge action and unimproved simple staggered fermion action  
were used in~\cite{Kogut:2015zta} with known large cutoff effects and lacking systematic control 
on the estimate of the renormalized coupling and its continuum limit, limited to simulations away from the
required chiral limit~\cite{Kogut:2015zta}.
In contrast, our direct determination of the $\beta$-function from the gauge field gradient flow method when
the cutoff is removed and the continuum
limit is taken shows agreement with the 2-loop $\beta$-function in the ${\rm g^2_R \sim 1-3}$ range 
without any sign of a conformal IRFP~\cite{Fodor:2015zna}. 
To resolve the apparent controversy, definitive and systematic finite temperature studies of ${\rm \chi SB}$  
would be needed in the massless fermion limit of this important gauge field theory
close to the lower edge of the conformal window.

\section{Electroweak multiplets and anomaly constraints}
The first direct test of the sextet theory is expected to come from the strongly coupled sector 
of the new gauge force which predicts 
resonances in the 2-3 TeV range within the reach of Run 2 at the LHC.
As an example, a rho-like vector state has been predicted in the model
at approximately 2 TeV which could be observed as a diboson resonance 
excess above LHC background events~\cite{Fodor:2012ty,Fodor:2012ni,Fodor:2014pqa,Fodor:2015vwa}.
In Section 7 we will briefly comment on the recently reported diboson excess
from the ATLAS and CMS collaborations, consistent with our prediction but far from settled.
The location of the rho-like resonance at 2 TeV would be less surprising in Higgsless Technicolor, 
but the emergent light ${\rm 0^{++}}$ scalar on the Electroweak scale, far 
separated from the resonance spectrum in sextet dynamics, is a distinct and unexpected new feature.
The recently found diphoton excess from
ATLAS and CMS resonance searches~\cite{diphoton:2015}  in the invariant 
mass range around 750 GeV would require very specific explanation
of a light flavor singlet near-conformal state separated from the 2-3 TeV composite resonance range
in strongly coupled  composite gauge theories~\cite{Molinaro:2015cwg}.

Building a BSM theory requires the embedding of the strongly coupled sextet fermion doublet
into the ${\rm SU(2)_w\!\otimes\!U(1)_Y}$ Electroweak gauge group with a new outlook 
beyond the Standard Model under global and gauge anomaly constraints. 
We will show that the general construction can accommodate new physics
on the Energy Frontier including new heavy leptons and massive neutrinos. The related dark
matter content of the theory implies interesting scenarios for future investigations. 
As a first step, model building requires a consistent 
Electroweak multiplet structure with a simple realization 
of the composite Higgs mechanism from sextet gauge dynamics.

\subsection{The Electroweak (EW) multiplet structure of sextet fermions}
As in the minimal scheme of Susskind~\cite{Susskind:1978ms} and Weinberg~\cite{Weinberg:1979bn}, 
the gauge group of the theory  is 
${\rm SU(3)_{bsm}\!\otimes\!SU(3)_c\!\otimes\!SU(2)_w\!\otimes\!U(1)_Y}$ where  ${\rm SU(3)_c}$ designates 
the QCD color gauge group and ${\rm SU(3)_{bsm}}$ represents the BSM color gauge group 
of the new strong gauge force.
In addition to quarks and leptons of the Standard Model, we include one ${\rm SU(2)}$ 
bsm-flavor doublet ${\rm  (u,d)}$ of fermions which are ${\rm SU(3)_c}$ singlets and transform 
in the six-dimensional sextet representation of bsm-color, distinct from
the fundamental color representation of fermions in the original Technicolor
scheme~\cite{Susskind:1978ms,Weinberg:1979bn}. 
The formal designation ${\rm  (u,d)}$ for the bsm-flavor doublet of sextet fermions uses a similar notation 
to the two light quarks of QCD but describes completely different physics. 
The massless sextet fermions form two chiral doublets 
${\rm (u,d)_L}$ and ${\rm (u,d)_R}$ under the global symmetry group
${\rm SU(2)_L\!\otimes\!SU(2)_R\otimes\!U(1)_B}$. Baryon number
is conserved 
for quarks of the Standard Model separate from baryon number conservation 
for sextet fermions which carry $1/3$ of BSM baryon charge associated with the BSM sector of the global ${\rm U(1)_B}$  symmetry.
	
It is straightforward to define 
consistent multiplets  for the sextet fermion flavor doublet 
under the ${\rm SU(2)_w\!\otimes\!U(1)_Y}$ Electroweak gauge group
with hypercharge assignments for left- and right-handed fermions 
transforming under the ${\rm SU(2)_w}$  weak isospin group.
The two fermion flavors ${\rm {u^{ab}}}$ and ${\rm d^{ab}}$ of the strongly coupled sector
carry six colors in two-index symmetric tensor notation,
${\rm a,b = 1,2,3}$,  associated with the gauge force of the ${\rm SU(3)_{bsm}}$ group.
This is equivalent to a six-dimensional vector notation in the sextet representation.

The fermions transform as left-handed weak isospin doublets
and right-handed weak isospin singlets for each color,
\begin{equation}
{\rm \psi^{ab}_L = \lmatrix{c}  {\rm u^{ab}_L} \\  {\rm d^{ab}_L }\rmatrix , \qquad \psi^{ab}_R = (u^{ab}_R ,\; d^{ab}_R) }.
\end{equation}
With this choice of representations, the normalization for the hypercharge ${\rm Y}$ 
of the ${\rm U(1)_Y}$ gauge group is 
defined by the relation ${\rm Y=2(Q-T_3)}$, with  ${\rm T_3}$ designating the third component of weak isospin.

Once Electroweak gauge interactions are turned on, 
the chiral symmetry breaking pattern  ${\rm SU(2)_L\!\otimes\!SU(2)_R \rightarrow SU(2)_V}$ 
of strong dynamics breaks  Electroweak symmetry in the expected pattern,
${\rm SU(2)_w \times U(1)_Y \rightarrow U(1)_{\rm em}}$, and with the simultaneous
dynamical realization of the composite Higgs mechanism. It is important to note that the 
dynamical Higgs mechanism is facilitated through the electroweak gauge couplings of
the sextet fermions and does not depend on the hypercharge assignments 
of the multiplets~\cite{Susskind:1978ms}. 

Hypercharges of left-handed doublets and right-handed singlets
are determined from anomaly constraints and  consistent
electric charge assignments for fermions.  
Apparently, there are two simple solutions to the anomaly constraints
with different hypercharge assignments for 
left-handed doublets. However,
only one of them avoids consistency problems with what we know about the relic abundance
of dark baryons from their primordial evolution in the early Universe. 
Before we describe the two different solutions to anomaly constraints in Section 4 with related implications,
it is useful to briefly review first the merits of the sextet model for the minimal 
realization of the composite Higgs mechanism
which is independent of the two different hypercharge choices for left-handed doublets.

\subsection{The minimal composite Higgs in the sextet model}
The chiral symmetry breaking pattern ${\rm SU(2)_L\!\otimes\! SU(2)_R\rightarrow SU(2)_V}$  of the 
${\rm SU(2)}$ BSM flavor group of sextet fermions generates an isotriplet of three massless Goldstone 
bosons in the chiral limit. 
The three Goldstones
will become longitudinal modes of the ${ W^{\pm}}$ and $Z^0$ weak gauge bosons via the dynamical 
Higgs mechanism when the Electroweak interactions are turned on. This minimal realization of the composite Higgs
mechanism comes from the perfect match between the longitudinal Electroweak gauge bosons 
and three massless Goldstone bosons  from sextet strong dynamics as one of the most 
attractive features of this BSM theory.
It is near-conformal with just one ${\rm SU(2)}$  fermion flavor doublet with unexpected spectroscopy,
distinct from old Technicolor. We already noted that the
dynamical Higgs mechanism does not depend on the electric charge assignment of the (u,d) fermion pair
with left-handed doublets and right-handed singlets as would be determined 
from the choice of hypercharges ${\rm Y}$
under the weak isospin gauge group ${\rm SU(2)_w}$. Independent of ${\rm Y}$, 
the three Goldstone bosons always have the correct 
integer electric charges ${\rm (\pm 1,0)}$ to morph into the longitudinal components of the weak bosons as further
detailed in Section 4.
Since the sextet gauge model is naturally located very close to the lower edge of the conformal window 
without fine tuning, a light scalar with a far separated resonance spectrum is expected from strong dynamics,
making the minimal Higgs mechanism of the model economic and attractive.

In contrast, the condition of near-conformal behavior with fermions  in the fundamental 
representations of ${\rm SU(3) }$ bsm-color
requires a large number of fermion flavors in the BSM construction which leads to an excess of unwanted 
massless Goldstone bosons. 
%
The added complexities can be illustrated with the well-known one-family model
from the Technicolor era with global flavor symmetry 
${\rm SU(8)_L\!\otimes\!SU(8)_R}$  of eight massless chiral fermions carrying bsm-color 
in the fundamental representation of the ${\rm SU(3)}$ gauge group~\cite{Farhi:1980xs}.
In the new near-conformal BSM paradigm a minimum of eight flavors is needed in the fundamental color
representation to get closer to the conformal window.  This motivated recent studies of the 
eight-flavor fermion model to understand the
strong BSM force of the one-family model from several lattice studies~\cite{Fodor:2009wk,Aoki:2014oha,Fodor:2015baa,Hasenfratz:2014rna,Appelquist:2014zsa,Schaich:2015psa}.
The  ${\rm \chi SB}$ pattern  ${\rm SU(8)_L\!\otimes\!SU(8)_R\rightarrow SU(8)_V}$  generates 63 Goldstone bosons 
with only three needed in the composite Higgs mechanism.
Generating masses for the excess unwanted Goldstone bosons  presents 
non-trivial phenomenological challenges, making it more difficult to achieve the 
desired goal of a near-conformal spectrum with a light $0^{++}$ scalar 
far separated from resonance excitations.

\section{Constraints from global and gauge anomalies}
Assuming that the strongly coupled gauge sector of the sextet model turns out to be compatible 
with resonances 
in the 2-3 TeV region, a more complete BSM outlook of the model would come into focus,
guided by anomaly constraints. 
We will show that the existence of stable baryons in the sextet model when combined with our understanding 
of the early Universe requires anomalous fermion representation in the Electroweak sector. 
Although the sextet BSM model cannot offer a UV complete solution
for fermion mass generation and the related flavor problem, the outlook for new physics 
from extending the strongly coupled gauge sector guided by anomaly constraints is necessary. 
Anomaly compensating new physics has several important and immediate aspects 
without deference to physics on the scale of UV completion. 
It provides an outlook and framework for new 
physics the model can predict, or accommodate on several energy scales. 
Anomalies not only can predict or accommodate plausible new
fermion content at the TeV scale but also give insight into
Electroweak corrections to Standard Model expectations.
The example we will provide below is the anomaly canceling 
pair of two massive lepton doublets (and associated right-handed singlets) to cancel the 
anomalies in Eqs.~(\ref{eq:Banomaly1}) and (\ref{eq:Banomaly2}) from left-handed sextet fermion
doublets. New charged lepton 
and neutrino masses are partially constrained parameters in this anomaly matching extension.
Even if the new fermion masses are set to very high energy scales, 
their infrared effects from the associated anomaly content will always survive. 
The best known examples of this footprint from integrating out heavy fermions include
the Wess-Zumino effective action and other residual effects on the Electroweak scale~\cite{D'Hoker:1984ph,D'Hoker:1984pc,Preskill:1990fr,D'Hoker:1992bv}.

\subsection{Anomaly conditions in the sextet model}
Anomaly constraints have a long history in Technicolor motivated BSM model building with 
representative examples 
in~\cite{Dietrich:2005jn,Kainulainen:2006wq,Foadi:2007ue,Antola:2009wq,Kainulainen:2009rb}.
The first condition for model construction with left-handed doublets is the global Witten anomaly 
constraint which requires an even number of left-handed SU(2) multiplets to 
avoid inconsistency in the theory from a vanishing fermion determinant of the partition function~\cite{Witten:1982fp}.

In addition, gauge anomaly constraints also have to be satisfied~\cite{Adler:1969er}. 
With vector current ${\rm  V^i_\mu(x)=\overline{\psi}T^i\gamma_\mu\psi(x)}$ and axial current 
 ${\rm A^i_\mu(x)=\overline{\psi}T^i\gamma_\mu\gamma_5\psi(x)}$
constructed from fermion fields and internal symmetry matrices ${\rm T^i}$ in some group representation 
R for fermions, the anomaly in the axial vector Ward
identity is proportional to ${\rm tr(\{T^i(R),T^j(R)\}T^k(R))}$ and must vanish. 
In the sextet theory fermions are either left-handed doublets or right-handed singlets 
under the ${\rm SU(2)_w}$ gauge group. 
The matrices ${\rm T^i}$  will be either the ${\rm \tau^i}$ Pauli matrices 
or the diagonal ${\rm U(1)}$ hypercharge ${\rm Y}$.
Since the ${\rm SU(2)}$  group is  anomaly free, ${\rm tr(\{\tau^i,\tau^j\}\tau^k)=0}$, 
we only need to consider anomalies where at least one ${\rm T^i}$ is the hypercharge Y.
The non-trivial constraints come from two conditions on hypercharge traces,

\begin{equation}
{\rm tr(Y )=0, \quad tr(Y^3) \propto tr(Q^2T_3 - QT^2_3) = 0}\;, 
\label{eq:hyper}
\end{equation}
where  ${\rm Y=2(Q-T_3)}$ with electric charge ${\rm Q}$, and 
${\rm T_3}$ as the third component of weak isospin. 

There are two simple solutions for BSM model building with sextet fermions to satisfy 
the Witten anomaly condition and
gauge anomaly constraints on tr(Y) and ${\rm tr(Y^3)}$ in Eq.~(\ref{eq:hyper}). 
The first solution with the choice ${\rm Y(f_L)=0}$ for doublets of left-handed sextet fermions ${(\rm f_L)}$
leads  to half-integer electric charges for composite baryons. The second solution
with the  choice ${\rm Y(f_L)=1/3}$ for doublets of left-handed sextet fermions leads to integer electric charges
for composite baryons.  The hypercharges of right-handed singlets are automatically set from
consistent electric charge assignments in both cases. The two choices are discussed next 
with their implications.

\subsection{EW content from Y(${\rm f_L}$)=0 with baryons of half-integer electric charge Q}
Since left-handed fermion doublets occur with an even number of sextet colors in the strong sector, 
the global anomaly condition is automatically satisfied for the first solution without  adding any new 
left-handed lepton doublets to the theory.
Gauge anomaly cancellation in this case requires Y=0  assigned 
to left-handed doublets of fermions 
with sextet color (a,b) in the two-index symmetric tensor representation,
\begin{equation}
{\rm Y(u_L^{ab}) =0, \quad Y(d_L^{ab}) = 0, \quad Y(u_R^{ab}) = 1, \quad Y(d_R^{ab}) = -1}\;,
\label{eq:typeA}
\end{equation}
leading to fractional charges ${\rm Q(u_L) = 1/2}$ and ${\rm Q(d_L) = -1/2}$ from the 
${\rm Y=2(Q-T_3)}$ relation. 
Hypercharges  in Eq.~(\ref{eq:typeA})  
are set for right-handed fermions  from consistent electric charge 
assignments ${\rm Q(u_R) = 1/2}$ and ${\rm Q(d_R) = -1/2}$. 

The minimal Electroweak content with the choice ${\rm Y(f_L)=0}$  
leads 
to baryon states of three constituents forming flavor isospin doublets 
with a half unit of positive electric charge for isospin +1/2 with (uud) content and a half unit of negative electric charge
for isospin -1/2 with (udd) content. This is in sharp contrast to electric charges carried 
by the proton and neutron in QCD
where the Witten anomaly of three left-handed color doublets of (u,d)  quarks 
is compensated by the left-handed lepton doublet of the Electroweak theory. 
The first generation of quarks and leptons then allows 
the well-known choice of electric charges  
${\rm Q(u) = 2/3}$ and ${\rm Q(d) = -1/3}$,
compatible with gauge anomaly constraints and the pattern applied to all three generations. 
As a consequence, baryons in QCD carry integer electric charges.
The unique solution to anomaly constraints in the Standard Model is consistent
with direct observations of the full
particle content including quark and lepton quantum numbers 
matching all the anomaly conditions.

%
In the sextet BSM theory we do not have direct observations on new heavy baryons
to set unique  hypercharge
assignments for left-handed doublets and right-handed singlets of sextet fermions from one of two alternate 
solutions to the anomaly conditions. 
Viability of the choices  ${\rm Y(f_L)=0}$, or  ${\rm Y(f_L)=1/3}$,  is
affected by the different electric charge assignments they imply. 
With heavy baryon masses in the 3 TeV range, as determined from our lattice simulations in Sections 6 and 7 outside the reach of immediate accelerator search, 
our understanding of the early Universe
provides important input concerning the two simple anomaly solutions.
The seemingly minimal solution with ${\rm Y=0}$ for left-handed doublets would lead to intriguing 
predictions of baryon states with half-integer electric charges for future accelerator searches and relics
with fractional electric charges from the early Universe with observable consequences.
Problems with half-integer electric charges, from the choice ${\rm Y(f_L)=0}$ in our case, were anticipated earlier 
from strong observational limits on stable fractional charges in the early Universe 
and their terrestrial relics~\cite{Chivukula:1989qb,Langacker:2011db}.
Specifically, the sextet model inheriting this problem (with a related discussion deferred to Section 5)
leads to the non-controversial ${\rm Y(f_L)=1/3}$ anomaly solution with new BSM implications and outlook.

\subsection{EW content from Y(${\rm f_L}$)=1/3 with baryons of integer electric charge Q}
Motivated by problems of the anomaly-free selection with half-integer electric charges,
we are now lead to consider the Electroweak content with sextet model baryons carrying integer electric charges 
which requires non-zero hypercharge for  left-handed fermions with sextet color, in close analogy
with the Standard Model pattern of fractional electric charges carried by three colors of quarks
and integer charges carried by baryons in QCD. 
Hypercharge assignment ${\rm Y=1/3}$ is set for the left-handed sextet fermion doublets with consistent
choices required for right-handed singlets,
\begin{equation}
{\rm Y(u_L^{ab}) = 1/3, \quad Y(d_L^{ab}) = 1/3, \quad Y(u_R^{ab}) = 4/3, \quad Y(d_R^{ab}) = -2/3}\;.
\label{eq:typeB}
\end{equation}
Eq.~(\ref{eq:typeB}) leads to QCD-like electric charge assignments for sextet fermions with ${\rm Q(u)=2/3}$ and ${\rm Q(d)=-1/3}$.
The ${\rm Y=1/3}$  hypercharge assignment for left-handed doublets
implies integer electric charges for composite baryons, built from three
fermions of sextet color. The fermion content of the baryon doublet is given by
${\rm (uud) ~isospin=+1/2, ~Q=+1}$   and ${\rm (udd) ~isospin=-1/2, ~Q=0}$,  in contrast to the
hypercharge selection ${\rm Y=0}$ leading to baryons of half-integer electric charges. 

With six left-handed doublets there is no Witten anomaly, but  ${\rm Y=1/3}$ for the left-handed doublets,
necessary to get integer electric charges 
for composite baryons, leads to gauge anomalies,
\begin{eqnarray}
&&{\rm tr(Y )=6\Big\{\frac{1}{3}\times 2 + \frac{4}{3} - \frac{2}{3}\Big\} = 8},  
 \label{eq:Banomaly1} \\ 
&&{\rm tr\big(Y^3\big) \propto tr\big(Q^2T_3 - QT^2_3\big)=
	6\Big\{\big(\frac{2}{3}\big)^2\times \frac{1}{2} - \big(\frac{1}{3}\big)^2\times \frac{1}{2} -\frac{2}{3}\times \frac{1}{4} 
	+ \frac{1}{3}\times \frac{1}{4} \Big\}=\frac{1}{2}}.     
\label{eq:Banomaly2}
\end{eqnarray}
The anomalies in Eqs.~(\ref{eq:Banomaly1}) and (\ref{eq:Banomaly2}) have 
to be compensated with new physics from some unknown scale. Once a commitment is made 
to the ${\rm Y=1/3}$ choice with integer electric charges for sextet model baryons,  infrared effects 
from anomaly compensating new physics cannot be ignored and will affect Electroweak precision tests
and other predictions on the Electroweak scale. 
As we will argue in Section 5, the non-controversial second type of  hypercharge choice, ${\rm Y(f_L)=1/3}$,
leads to relic stable baryons from the early Universe. These baryons are neutral
without direct conflict from limits on galactic and terrestrial charged relics but they require  new physics
to compensate the  anomalies in Eqs.~(\ref{eq:Banomaly1}) and (\ref{eq:Banomaly2}).

\subsection{Y(${\rm f_L}$)=1/3 anomaly cancellation with new left-handed lepton doublets}
The viable extension of the strongly coupled sextet gauge sector is anticipated from the anomaly 
cancellation mechanism of new left-handed fermion doublets. 
The absence of a global SU(2) anomaly in the strongly coupled gauge sector requires the addition of a 
pair of left-handed fermion doublets. They are introduced as singlets under SU(3) QCD color
and SU(3) bsm-color. 
Gauge anomaly constraints require the choice  ${\rm Y\neq 0}$ hypercharge assignment 
for the left-handed fermion doublets to compensate the anomalies from sextet fermions as counted in 
Eqs.~(\ref{eq:Banomaly1}) and (\ref{eq:Banomaly2}). Consistent hypercharge 
assignments for right-handed singlets completes the solution for anomaly cancellation.
For simplicity,
we will use the notation of lepton families with a new family label   ${\rm \alpha = 1,2}$,
\begin{equation}
{\rm \lmatrix{c} {\rm N^{(\alpha)}_L }\\ {\rm E^{(\alpha)}_L} \rmatrix , \quad  (N^{(\alpha)}_R ,\; E^{(\alpha)}_R),  \quad \alpha=1,2 }.
\end{equation}
It should be noted that the addition of the two lepton families is different from adding complete 
generations of quarks and leptons in the Standard Model. Here lepton families are added only to the
sextet fermion doublet of the strongly coupled gauge sector. 

As described before, the hypercharge assignment ${\rm Y=1/3}$ is set for the left-handed 
sextet fermion doublet with ${\rm Y=4/3}$ for the right-handed singlet ${\rm u_R}$ and
${\rm Y=-2/3}$ for ${\rm d_R}$,
leading to QCD-like charge assignments for sextet fermions with ${\rm Q(u)=2/3}$ and ${\rm Q(d)=-1/3}$.
The added pair of lepton doublets, without introducing global anomalies, allows  QCD-like charge assignment 
for the sextet fermion doublet by canceling the gauge anomalies when hypercharge ${\rm Y=-1}$ 
is set for  the new left handed leptons ${\rm N_L^{(\alpha)}}$, ${\rm E_L^{(\alpha)}}$.
The right-handed singlets  ${\rm N_R^{(\alpha)}}$ are assigned ${\rm Y=0}$, and 
${\rm Y=-2}$ is set for the right-handed singlets ${\rm E_R^{(\alpha)}}$. It is easy to check that both gauge 
anomaly constraints of Eq.~(\ref{eq:hyper}) are satisfied. 
The new leptons ${\rm E^{(\alpha)}}$ carry electric charge ${\rm Q=-1}$
and electric charge Q=0 is set for the massive neutrinos ${\rm N^{(\alpha)}}$. 

The Lagrangian of the two lepton flavors include gauge invariant mass terms for charged 
leptons and massive neutrinos~\cite{Kainulainen:2009rb,Akhmedov:2014kxa}. 
The most general mass matrix
with mixing describes Dirac masses for 
the charged leptons ${\rm E_-^{(\alpha)}}$ with two additional terms
representing Majorana masses for ${\rm N_L^{(\alpha)}}$ and ${\rm N_R^{(\alpha)}}$.
After diagonalization of the ${\rm 2\times 2}$ mass matrix, each lepton family
will have two
neutrino mass eigenstates ${\rm M_1}$ and ${\rm M_2}$ in addition to the charged lepton mass ${\rm M_E}$ and
a tunable mixing angle $\Theta$ from the Lagrangian mass parameters~\cite{Akhmedov:2014kxa}.
The two families can be allowed to mix which leads to more options in the full mass spectrum.

A more comprehensive analysis of the lepton sector in the sextet BSM model is beyond the scope
of this work and will be reported in a separate publication~\cite{Kuti:2015}. Here the lepton sector 
serves to illustrate the most
plausible framework for BSM extension of the sextet model, compatible with anomaly conditions 
and integer electric charges for baryons. The lepton sector with its spectrum and quantum numbers 
also provides useful general guidance for expected new effects on the Electroweak scale from a broad 
range of mass parameters, or equivalently the low energy effective action with a Wess-Zumino term and
other residual anomaly effects if we seek to integrate out the leptons 
asymptotically in the heavy mass limit. 
Clearly, the BSM outlook of the sextet model remains quite flexible and interesting. 
Here we only briefly summarize our main findings so far:

\begin{itemize}
	\item There exists a range of charged lepton masses and heavy neutrino masses which are
	not in conflict with Electroweak precision tests. In addition, in that range the charged leptons
	can decay and the lowest stable mass is a Majorana neutrino.	
	\item Stable and heavy Majorana neutrinos are  interesting Dark Matter candidates in the
	mass range where their relic densities are not	in conflict with direct dark matter experiments, like 
	XENON100~\cite{Aprile:2012nq} and LUX2013~\cite{Akerib:2013tjd}.
	\item The Wess-Zumino action and its effects on the Electroweak scale can be identified from the 
	footprints of  heavy leptons and neutrinos of the model at very high energy scales.
	\item Restrictions on heavy fermions from vacuum instability of the effective potential with 
	the composite  Higgs remain a difficult and  unresolved problem.
\end{itemize}

Leaving further analysis of the lepton sector for future reporting, we will turn in Sections 6 and 7 
to the baryon spectrum
from non-perturbative simulations since it is not affected by the intriguing properties of the lepton sector.
In Section 5  we will explain first in some detail why fractional electric charge assignment for baryons
is problematic and most likely excluded by what we understand from the early history of the Universe. 
%
The well motivated ${\rm Y(f_L)=1/3}$ solution to the anomaly constraints with two lepton flavors 
leads to stable relic baryons from the early Universe. These baryons are neutral
and contribute to the missing dark matter content of the Universe. 
Their relic abundance and direct detection limits in dark matter experiments 
require quantitative analysis decoupled from future developments in the lepton sector.

\section{Sextet model baryons and the early Universe}

There is an 
exactly conserved ${\rm U(1)}$ symmetry which, when combined with exact electric charge conservation, 
means that there is a lightest stable baryon state in the spectrum which will be the primary focus of 
the forthcoming discussion. There are several questions to consider:

\hskip 0.2 in -- the electric charge of the lightest and stable sextet model baryon,
\vskip 0.02 in 
\hskip 0.2 in -- their galactic and terrestrial relic abundance,
\vskip 0.02 in 
\hskip 0.2 in -- limits from direct detection in dark matter experiments.

\noindent We will discuss these questions with two different choices of hypercharges for the left-handed 
doublets of sextet fermions.

\subsection{Sextet model baryons from Y(${\rm f_L}$)=0  with half unit of electric charge}

The choice ${\rm Y(f_L)=0}$ for left-handed bsm-flavor doublets of fermions 
in the sextet bsm-color representation 
leads to composite baryon states in the 3 TeV mass range with spin one-half and electric charge ${\rm Q=\pm 1/2}$.
The two lightest baryon states form a degenerate pair and transform as an isospin doublet  ${\rm (uud,udd)}$ of bsm-flavor carrying half-integer electric charges of opposite sign. 
Since the lightest sextet baryon carries half-integer charge it
remains stable after its formation in the early Universe below the Electroweak transition temperature.
Additional speculations on some charge conservation violating mechanism to make fractionally charged baryons
unstable are outside the scope of the model and our discussion.
The lowest stable baryon state of the sextet model with spin one-half and electric charge one-half 
under this anomaly free scenario belongs to the class of fractionally charged massive particles (FCHAMP)
which have been discussed in several aspects before~\cite{DeRujula:1989fe,Langacker:2011db}. 

Arguments were presented against fractionally charged leptons with detailed estimates on their relic terrestrial density 
from the early Universe, strongly violating observational limits~\cite{Perl:2009zz,Langacker:2011db}. 
Estimates of the relic terrestrial density
of sextet model baryons with half-integer charge proceed along similar lines with some  uncertainties 
from non-perturbative strong gauge dynamics 
binding the fermions into baryons. We will briefly review the charge-symmetric evolution of 
these baryons in the early Universe. 
Some assumptions we will make on annihilation cross sections from strong gauge dynamics 
are unlikely to affect the qualitative conclusions on this problematic
anomaly-free choice. Charge-asymmetric evolution would make the scenario even less likely.


In the symmetric thermal evolution under discussion, baryons and antibaryons will remain  in
thermal equilibrium with decreasing charge-symmetric densities well below the Electroweak transition temperature.
At some freeze-out temperature ${T_*}$ the annihilation rate of baryons and antibaryons 
cannot keep up any longer with the expansion rate of the Universe. The 
total number of baryons and antibaryons remains approximately constant after freeze-out
for ${\rm T \ll T_*}$ and the relic abundance level is set
from the solution of the Boltzman equation~\cite{Steigman:1979kw}. The freeze-out temperature  
and the relic sextet baryon number density ${\rm n_{B_6}}$ relative to ordinary baryon number density ${\rm n_B}$
will depend on the sextet baryon mass ${\rm M_{B_6}}$ mass and the thermally averaged annihilation rate 
${\rm \langle\sigma v \rangle_{ann}}$ of sextet model baryons and antibaryons,
\begin{equation}
{\rm  \frac{n_{B_6}}{n_B} \approx \frac{10^{-25}}{M_{B_6}\langle\sigma v \rangle_{ann}}}\;.
\end{equation}
\label{eq:annihil}
The velocity dependent annihilation cross section ${\rm \sigma}$ is known for ordinary nucleons
and will be estimated for sextet model baryons whose mass is approximately ${\rm M_{B_6}=3~TeV}$, 
as reported in Section 7 from our non-perturbative 
lattice simulations. For a qualitative estimate, the thermally averaged annihilation rate 
of sextet model baryons is scaled down from the 
nucleon-antinucleon annihilation cross sections of QCD according to the generally accepted approximation,
\begin{equation}
{\rm  \langle v\sigma \rangle_{ann} \approx  \langle v\sigma \rangle^{nuc}_{ann}\times M^2_{nuc}/M^2_{B_6} }\;. 
\label{annihil6}
\end{equation}

Based on the value of ${\rm M_{B_6}}$ we determined and using the rough estimate of the annihilation cross section,
the freeze-out temperature and the ratio ${\rm x=M_{B_6}/T_* }$  can be approximately determined, 
with the ratio logarithmically dependent on the thermally averaged annihilation cross section. 
This  leads to an approximate relic sextet baryon number density as a fraction of nucleon number density,
\begin{equation}
{\rm  \frac{n_{B_6}}{n_B} \approx 3\cdot 10^{-7}},
\label{eq:annihil6}
\end{equation}
far exceeding terrestrial limits of stable fractional charges. The factor $3$ in Eq.~\ref{eq:annihil6} is associated 
with the particular assumption about the annihilation rate and the details of the freeze-out calculation. 
Only the order of magnitude estimate is relevant for the argument in what will follow. 
As pointed out in~\cite{Langacker:2011db}, fractionally charged baryons and anti-baryons
will get re-thermalized  at ${\rm \approx ~ 300~K}$  on Earth 
and continue annihilating over the 4.5 Gyr life of Earth.
This, at first thought, perhaps would bring their terrestrial density below acceptable
observational limits, many orders of magnitude less than the ${\rm 3\cdot 10^{-7} }$ freeze-out relic abundance.
Unfortunately the terrestrial annihilation mechanism is blocked by some overlooked new mechanism
in the early Universe where negatively charged sextet model baryons will capture alpha particles 
with calculable estimates of the capture rate~\cite{Langacker:2011db} and significant relic density compound
particles ${\rm( \alpha^{++}B^-)}$. 
This will block the terrestrial annihilation for a large fraction of the positively charged free baryons with negatively
charged bound baryons which are screened by alpha particles of the compound,
hiding behind a repulsive Gamow barrier.
As noted in~\cite{Langacker:2011db} the terrestrial annihilation is unlikely to continue at the necessary rate
for fractionally charged leptons because 
negative charges will bind to alpha particles with calculable rate estimates with the Gamow barrier
blocking annihilation.   Similarly, relic positively charged
baryons cannot continue terrestrial annihilation at the needed rate, 
blocked by the repulsive Coulomb barrier between unbound  positively charged relic  baryons and 
the compound ${\rm( \alpha^{++}B^-)}$ objects, 
so that the terrestrial bounds 
most likely remain in violation. Other difficulties were also noted, like the symmetric 
distribution of opposite sign fractional charges in the interstellar medium of the galaxy which is also problematic  
for detection~\cite{Langacker:2011db}.

Unless unforeseen considerations bring new arguments for the viability of stable sextet model baryons with half-unit of
electric charges, their existence from the early Universe makes the ${\rm Y(f_L)=0}$ anomaly-free choice 
very unlikely. We will discuss next the 
more realistic solution to the anomaly-free construction.

\subsection{Sextet model baryons from Y(${\rm f_L}$)=1/3 with integer units of electric charge}
The lightest baryons in the strongly coupled sextet gauge sector are expected to form 
isospin flavor doublets  ${\rm (uud,udd)}$, similar to the pattern in QCD.
As we noted earlier, baryons in the sextet model  
should carry integer multiples of electric charges if ${\rm Y(f_L)\neq 0}$ to avoid problems with the relics
of the early Universe. This leads to the
simplest choice  ${\rm Y(f_L)=1/3}$ with gauge anomalies to be compensated. 
A new pair of left-handed lepton 
doublets emerged from this choice in Section 4 as the simplest manifestation 
of the anomalies and the Electroweak extension of the strongly coupled sextet gauge sector.

Neutron-like ${\rm udd}$ sextet model baryons ${\rm (n_6)}$ will carry no electric charge 
and proton-like ${\rm uud}$ sextet model baryons ${\rm (p_6)}$ have one unit of positive electric charge 
from the choice ${\rm Y(f_L)=1/3}$. 
The two baryon masses are split by electromagnetic interactions.
The ordering of the two baryon masses in the chiral limit of massless sextet fermions will
require non-perturbative {\em ab initio} lattice calculations of the electromagnetic 
mass shifts to confirm intuitive expectations
that the neutron-like ${\rm n_6}$ baryon has lower mass than the proton-like ${\rm p_6}$ baryon.
In QCD this pattern was confirmed by recent lattice calculations~\cite{Borsanyi:2014jba}.
We expect the same ordering in the sextet model so that the proton-like ${\rm p_6}$  baryon 
will decay very fast, ${\rm p_6 \rightarrow n_6 + ... }$,  with a lifetime ${\rm \tau \ll 1~second}$. 
It is unlikely for rapidly decaying  ${\rm p_6}$ baryons to leave any relic footprints 
from dark nucleosynthesis before they decay.

With BSM baryon number conservation the neutral ${\rm n_6}$ baryon is stable 
and observational limits on its direct detection from experiments like 
XENON100~\cite{Aprile:2012nq} and LUX2013~\cite{Akerib:2013tjd} have to be estimated.
In charge symmetric thermal evolution sextet model baryons are produced with relic number density ratio
${\rm  n_{B_6}/n_B  \approx 3\cdot 10^{-7}}$ (Eq.~\ref{eq:annihil}).
For 3 TeV sextet model baryon masses we can estimate the detectable dark matter ratio of 
respective mass densities ${\rm \rho_{B_6}}$ and ${\rm \rho_{B}}$ as
${\rm \rho_{B_6}/\rho_B  \approx 10^{-4}}$,
about ${\rm 5\cdot 10^{4}}$ times less than the full amount of unaccounted dark mass, 
${\rm \rho_{dark} \approx 5\cdot \rho_B}$.  
We will use this mass density estimate to guide observational limits on relic sextet model baryons
emerging from charge symmetric thermal evolution where
tests of dark baryon detection come from elastic collisions with nuclei in dark matter detectors.
The neutral and stable ${\rm n_6}$ 
baryon can interact several different ways with heavy nuclei in direct detection experiments
including (a) magnetic dipole interaction, (b) Z-boson exchange, (c) Higgs boson exchange, and 
(d) electric polarizability.

A brief review of our estimates of these interactions will lead us to important observations from what follows.
It turns out that cross sections from (a) and (b) can be parametrized and well estimated without 
lattice simulations.

(a) The magnetic moment ${\rm \mu_6 = g\cdot e/2M_{n_6}}$  of the neutral ${\rm n_6}$ sextet baryon
can be calculated from first principles on the lattice but the only unknown quantity, g,
is not needed in our estimate. The magnetic moment ${\rm \mu_6}$ of ${\rm n_6}$ controls the coherent scattering cross section 
from magnetic moments of protons and neutrons in heavy nuclei with slow elastic recoil in direct detection 
experiments. If the mass density of relic ${\rm n_6}$ baryons would be large enough to match 
all the missing dark matter, a limit on the magnetic dipole g-factor from LUX2013 would be set 
to ${\rm g^2 \leq (M_{n_6}/5.1~TeV)^3}$, otherwise they would have been detected~\cite{Akerib:2013tjd}. 
With ${\rm M_{n_6} = 3~TeV}$ the limiting value ${\rm g=0.45}$ can be far exceeded 
due to the much lower relic mass density of ${\rm n_6}$ in the symmetric thermal evolution of the Universe. 
The precise value of ${\rm g}$ would require a straightforward lattice calculation which is less
important with low magnetic dipole cross sections well below observational limits in comparison with
cross section estimates from Z-exchange.

(b) The sextet model ${\rm n_6}$ baryon carries isospin 1/2 and hypercharge 
${\rm Y=1}$ which is the source to the Z-boson field.
Coherent Z exchange between the ${\rm n_6}$ baryon and heavy nuclei in detectors of direct searches
leads to larger cross section than magnetic dipole scattering and detectability has to be carefully
calculated and compared with XENON100 data.
Is the sextet model with its relic  baryon density  still safe against the 
most sensitive detection limits in charge symmetric thermal evolution?  This turns out to be the most sensitive
test of the model. For dark matter candidates at ${\rm M_{n_6} = 3~TeV}$ XENON100 
sets a cross section bound of approximately
${\rm 10^{-43} cm^2}$ per nucleon under the assumption of full dark matter missing mass density of
${\rm \rho_{dark} \approx 5\cdot \rho_B}$.
Now our estimate of the cross section from Z exchange per nucleon for  ${\rm n_6}$  is 
approximately ${\rm 10^{-39} cm^2}$, seemingly four orders of magnitude above detection threshold. 
This is not the case however, because the ${\rm n_6}$ relic charge symmetric mass density is about 
${\rm 5\cdot 10^{4}}$ times less than the full amount of unaccounted dark mass. This leads to the
interesting observation that the  ${\rm n_6}$ sextet model baryons might be detectable in the 
next generation of direct searches. There are several caveats to this including uncertainties in estimating
the relic density from hypothesized annihilation cross sections and the complications of asymmetric thermal
evolution. 

(c) The Higgs exchange effect is expected to be small but it will require lattice calculation to determine the 
coupling of the composite light scalar to the ${\rm n_6}$ sextet model baryon. We will return to this problem
in a future report.

(d) Similarly, the estimate of the scattering cross section from electric polarizability of ${\rm n_6}$ baryons requires 
lattice calculations which are left for future work. This effect is expected to be much smaller than cross sections
for (a) and (b).

Based on these estimates we conclude that the sextet BSM model is consistent with observational limits and 
will contribute a small fraction to the missing dark matter content.
As a last and important step of our analysis, we turn now to the non-perturbative lattice determination of the baryon masses.

\section{Construction of the sextet nucleon operator}

We next discuss how to build a sextet baryon operator that can be used in lattice simulations to isolate the baryon state and measure its mass. In the first two parts of this section we discuss the color, spin and flavor structure of the sextet baryon
state in the continuum.  We will see that a symmetric color contraction is needed in order to construct a color singlet three-fermion state when fermions are in the sextet representation of ${\rm SU(3)}$. This is opposite to the behavior in QCD, where the baryon color wave function is antisymmetric. Consequently, the  construction of the sextet baryon operator is non-trivial using staggered lattice fermions, which we describe in the third part of this section. 

\vspace{-0.1cm}

\subsection{Color structure}

\vspace{-0.1cm}

Three ${\rm SU(3)}$ sextet fermions can give rise to a color singlet. The tensor product $6 \otimes 6 \otimes 6$ can be decomposed 
into irreducible representations of ${\rm SU(3)}$ as \cite{maria}
\bea
6 \otimes 6 \otimes 6= 1\oplus2\times 8\oplus 10\oplus \ol{10}\oplus 3\times 27\oplus 28\oplus 2\times 35,
\eea
where irreps are denoted by their dimensions and $\ol{10}$ is the complex conjugate of 10.
The color singlet state corresponds to the unique singlet above. Fermions in the $6$-representation $\psi_{ab}$ are symmetric in the two indices and transform as
\bea
\psi_{aa'} \longrightarrow U_{ab}~U_{a'b'}~\psi_{bb'}
\eea 
and the color singlet combination is given by
\bea
\epsilon_{abc}~\epsilon_{a'b'c'}~\psi_{aa'}~ \psi_{bb'}~ \psi_{cc'}\;.
\eea 
(We earlier used superscripts for clarity.) Let us introduce the index $A=1,\ldots, 6$ for the $6$ components of the symmetric $\psi_{ab}$, i.e.~switch notation to
$\psi_{ab} = \Psi_A$. Then the above color singlet operator may be written as
\bea
\epsilon_{abc}~\epsilon_{a'b'c'}~\psi_{aa'}~ \psi_{bb'}~ \psi_{cc'} = T_{ABC}~ \Psi_A~ \Psi_B~ \Psi_C,
\label{colorsinglet}
\eea
with a completely symmetric 3-index tensor $T_{ABC}$. The contrast with QCD where the baryon color contraction is antisymmetric is here explicit.

\vspace{-0.1cm}

\subsection{Spin flavor structure}

\vspace{-0.1cm}

As we have seen the color contraction is symmetric for the sextet representation and hence the overall antisymmetry
of the baryon wave function with respect to the interchange of any two fermions  must come from the spin-flavor structure. Our operator construction is fully relativistic, we look here at the non-relativistic limit for illustration, omitting color indices. We label the two flavors 
$u$ and $d$ as in QCD and the non-relativistic spin will be either $\uparrow$ or $\downarrow$. We start with $| \uparrow u, \uparrow d, \downarrow u \rangle$ and build the desired state by requiring it to be antisymmetric under all possible interchanges, leading to
\bea
\label{psi}
| \uparrow \psi  \rangle = 
| \uparrow u, \uparrow d, \downarrow u \rangle + 
| \downarrow u, \uparrow u, \uparrow d \rangle + 
| \uparrow d, \downarrow u, \uparrow u \rangle - \nn \\
| \downarrow u, \uparrow d, \uparrow u \rangle - 
| \uparrow d, \uparrow u, \downarrow u \rangle - 
| \uparrow u, \downarrow u, \uparrow d \rangle\;,
\eea
which is similar to the wave function of triton~\cite{Derrick58}.

\vspace{-0.1cm}

\subsection{From continuum Dirac to lattice Staggered basis}

\vspace{-0.1cm}

We next convert from continuum to staggered lattice fermion operators. The lattice operators that create the state (\ref{psi}) belong to a suitable multiplet of ${\rm SU(4)}$ taste symmetry. Our staggered fermion operator construction follows \cite{gs, KlubergStern:1983dg, Altmeyer:1992dd, Ishizuka:1993mt}. We first convert from continuum operators to lattice operators in the Dirac basis, then we switch to lattice staggered fields. For simplicity we want to have operators as local as possible, thus in Dirac basis, our sextet baryon operator takes the form
\bea
N^{\alpha i}(2y)~=T_{ABC}~ u_A^{\alpha i}(2y) ~[u_B^{\beta j}(2y) ~(C \gamma_5 )_{\beta \gamma} 
~(C^* \gamma_5^*)_{jk}~d_C^{\gamma k}(2y)]
\eea
where Greek letters and lower case Latin letters denote spin and taste indices respectively. $C$ is the charge conjugation matrix satisfying
\bea
C\gamma_\mu C^{-1}=-\gamma_\mu^{{\rm T}},\nn\\
-C=C^{{\rm T}}=C^\dagger=C^{-1}.
\eea
The coordinate $y$ labels elementary staggered hypercubes.
Staggered fields are defined as
\bea
u^{\alpha i}(2y)=\frac{1}{8}\sum_\eta \Gamma^{\alpha i}_\eta~ \chi_u(2y+\eta)\;, \nn
\eea
where $\Gamma(\eta)$ is an element of the Euclidean Clifford algebra labeled by the four-vector 
$\eta$ whose elements are defined mod 2 as usual.
More precisely 
$\Gamma(\eta)=\gamma_1^{\eta_1}\gamma_2^{\eta_2}\gamma_3^{\eta_3}\gamma_4^{\eta_4}$ where 
$\eta \equiv (\eta_1, \eta_2, \eta_3, \eta_4)$.
Written in terms of the staggered fields,
\bea
N^{\alpha i}(2y)~=-T_{ABC}~\frac{1}{8^3}\sum_{\eta'} \Gamma^{\alpha i}_{\eta'}~ \chi_u^A(2y+\eta')
~\sum_{\eta }  S(\eta)\chi_u^B(2y+\eta) \chi_d^C(2y+\eta)\;,
\eea
where $S(\eta)$ is a sign factor.
To obtain a single time slice operator an extra term has to be either added to or subtracted from the diquark operator to cancel
the spread over two time slices of the unit hypercube. This is similar to what is done to construct 
the single time slice meson operators in QCD. This extra term corresponds to the parity partner of the nucleon.
The single time slice nucleon operator reads
\bea
N^{\alpha i}(2y)~=-T_{ABC}~\frac{1}{8^3}\sum_{{\vec \eta'}} \Gamma^{\alpha i}_{{\vec \eta'}}~ \chi_u^A(2y+{\vec \eta'})
~\sum_{{\vec \eta} }  S({\vec \eta})\chi_u^B(2y+{\vec \eta}) \chi_d^C(2y+{\vec \eta}).
\eea
This operator is a sum of $8\times 8=64$ terms over the elementary cube in a given time slice. 
The local terms vanish individually after the symmetric 
color contraction. The non-vanishing terms are those where a diquark resides on a corner of the cube at a fixed 
time-slice and the third fermion resides
on any of the other corners. The nucleon operator is thus the sum of $56$ such terms with appropriate sign factors. 
In order to find the mass of the lowest lying state any one of these $56$ terms can in principle be used.
We list in Table \ref{operators} the operators that we have implemented.

\begin{table}
\begin{center}
\begin{tabular}{|l|l|l|} 
\hline
Label  &  Operators (set a) & Operators (set b)  \\
\hline
 ${\rm IV_{{\rm xy}}}$   & $\chi_u(1,1,0,0)~\chi_u(0,0,0,0)~\chi_d(0,0,0,0)$&$\chi_u(0,0,0,0)~\chi_u(1,1,0,0)~\chi_d(1,1,0,0)$\\
 \hline
 ${\rm IV_{{\rm yz}}}$   & $\chi_u(0,1,1,0)~\chi_u(0,0,0,0)~\chi_d(0,0,0,0)$&$\chi_u(0,0,0,0)~\chi_u(0,1,1,0)~\chi_d(0,1,1,0)$\\
 \hline
 ${\rm IV_{{\rm zx}}}$   & $\chi_u(1,0,1,0)~\chi_u(0,0,0,0)~\chi_d(0,0,0,0)$&$\chi_u(0,0,0,0)~\chi_u(1,0,1,0)~\chi_d(1,0,1,0)$\\
 \hline
 VIII   & $\chi_u(1,1,1,0)~\chi_u(0,0,0,0)~\chi_d(0,0,0,0)$&$\chi_u(0,0,0,0)~\chi_u(1,1,1,0)~\chi_d(1,1,1,0)$\\
 \hline
\end{tabular}
\end{center}
\caption{The set of staggered lattice baryon operators we used to determine the baryon mass.}
\label{operators}
\end{table}  

\vspace{-0.3cm}

\section{Lattice simulations}

\vspace{-0.2cm}

We use the same lattice action as in our other studies of the sextet model, namely the tree-level Symanzik-improved gauge
action and the staggered fermion matrix with two stout steps of exponential smearing of the gauge link variables with two
stout steps~\cite{Morningstar:2003gk, Aoki:2005vt}. We implement the RHMC algorithm with the rooting procedure in all simulations to study the model with two fermion flavors. To accelerate the molecular dynamics time evolution we use multiple time scales~\cite{Urbach:2005ji} and the Omelyan integrator~\cite{Takaishi:2005tz}. The results we show here are at one lattice spacing corresponding to the bare gauge coupling $\beta = 6 / g^2 = 3.2$, which is defined as the overall prefactor of the Symanzik lattice action. We have continuing studies of the spectrum on finer lattice spacings to allow us to quantify lattice artifacts and determine the continuum limit of the spectrum, which we will report on in future publications. We examine the time histories of the correlators, the fermion condensate, the topological charge, and the gauge field energy on the gradient flow to estimate autocorrelation times. For the estimate of the statistical errors of hadron masses we used correlated fitting of the effective masses with a double jackknife procedure applied to the covariance matrices~\cite{delDebbio}.

\subsection{Nucleon operator comparison}

\begin{figure}
\begin{center}
\scalebox{0.75}{\includegraphics[angle=0,width=0.85\textwidth]{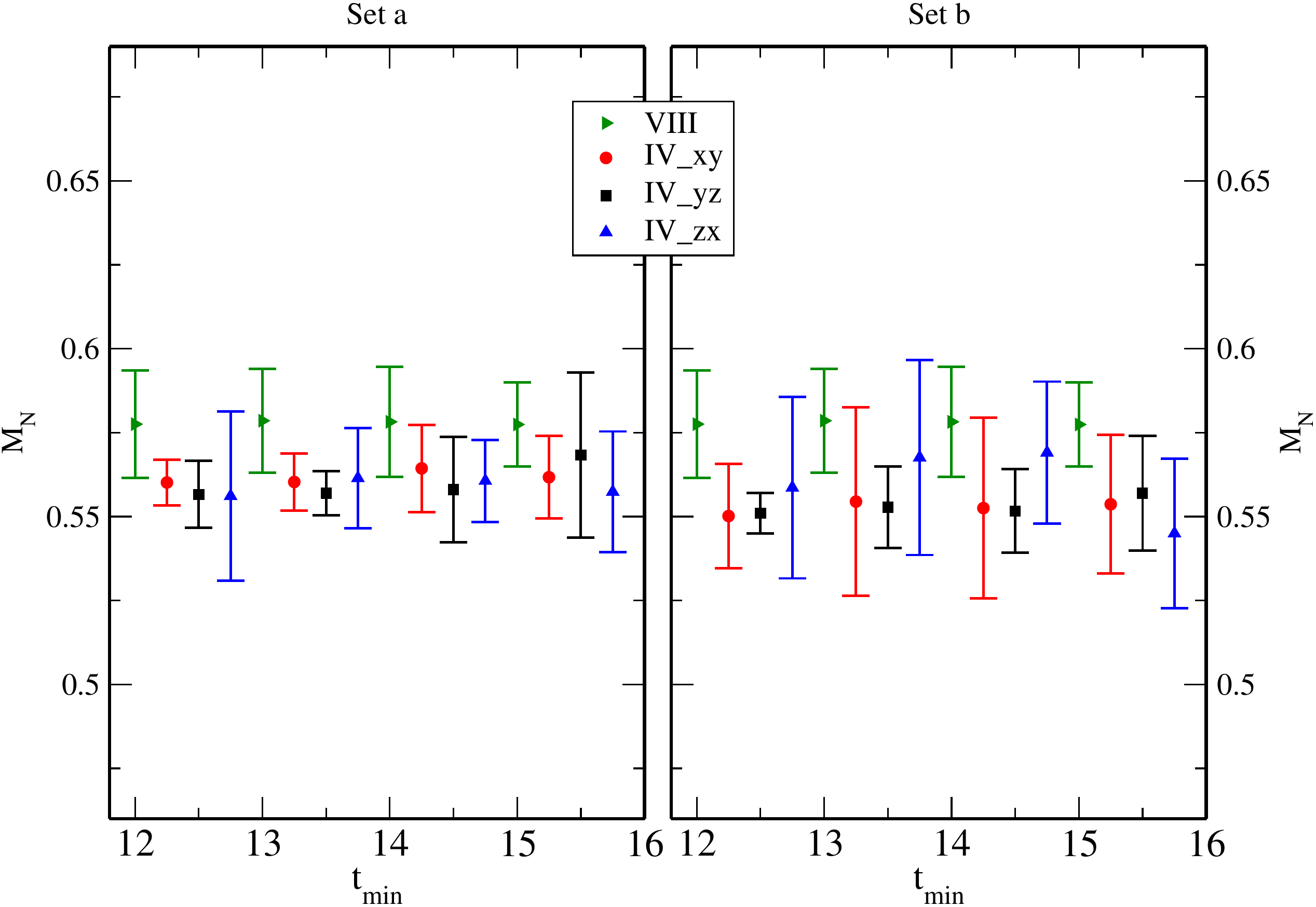}}
\vskip -.3 cm
\caption{{\footnotesize Comparison of $M_N$ from different operators varying $t_{min}$ 
with fixed $t_{max}=20$. The calculation is performed on lattices with $\beta=3.20$,
$V = 32^3 \times 64$ and $m=0.007$ over approximately 1000 trajectories. The $t_{min}$ 
values for the operators $IV_{xy}$, $IV_{yz}$ and $IV_{zx}$ are shifted by $0.25$,
$0.5$ and $0.75$ respectively for clarity. Sets a and b correspond to the location of the diquark operator as in Table~\ref{operators}.}}\label{signal_comparison}
\end{center}
\end{figure}

We investigate the quality of the signal for the operators listed in Table \ref{operators} on ensembles of approximately 1000 trajectories, each measurement separated by 50 trajectories during the molecular dynamics evolution, on a lattice volume $V=32^3 \times 64$ and at fermion mass $m=0.007$. For each operator the nucleon mass, $M_N$, is determined by correlated fitting of the effective mass with a double jackknife procedure applied to the covariance matrices from time separation $t_{min}$ to $t_{max}$. Figure \ref{signal_comparison} compares the
corresponding fits for various values of $t_{min}$ at $t_{max}=20$. It is observed that,
for all operators, the fits are not very sensitive to the choice of the fit range. Moreover, all
operators give consistent results for the nucleon mass within errors. The noise-to-signal
ratio is around $\sim 5\%$ for all operators, none of which is significantly
less noisy than the others. Therefore the quality of the resulting spectroscopy is
independent of the choice of operator, and in the following analysis we use the operator
$IV_{xy}$ from set $a$.

\subsection{First results}

\begin{figure}
\begin{center}
\scalebox{0.65}{\includegraphics[angle=0,width=1.0\textwidth]{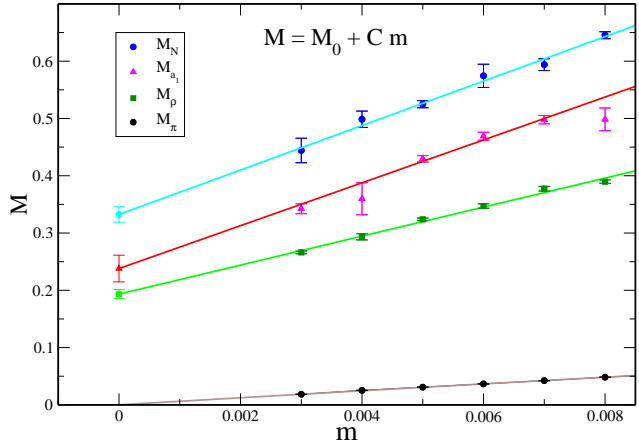}}
\vspace{-0.1cm}
\caption{
{\footnotesize Chiral extrapolation of $M_N$ (blue) in comparison with
$M_\pi$, $M_{a_1}$ and $M_\rho$. 
The calculation is performed on lattices with $\beta=3.20$, $V=48^3 \times 96$ 
for $m=0.003$ (except for $M_N$ on $V=32^3 \times 64$) and $V=32^3 \times 64$ for 
the heavier fermion masses, using 200 to 300 configurations at each mass.}
}
\label{chiral}
\end{center}
\end{figure}

In this section we present our first results for the nucleon spectroscopy. Simulations and data analysis of the sextet model are continuously ongoing, what we show here is a snapshot of the results at one lattice spacing. The nucleon correlator of operator
$IV_{xy}$ in set $a$ is measured on ensembles with lattice volume $V = 32^3 \times 64$ and fermion masses ranging from $m=0.003$ to $m=0.008$, each with 200 to 300 configurations, each configuration separated by 5 Monte Carlo trajectories. In Figure \ref{chiral} we show the chiral extrapolation of $M_N$, as well as extrapolations of the Goldstone boson, the $a_1$ and $\rho$ mesons, denoted by $M_\pi$, $M_{a_1}$ and $M_\rho$ respectively \cite{Fodor:2012ty, ourfuture}. For all except the Goldstone boson state we assume linear dependence on the fermion mass towards the chiral limit. We see the baryon remains significantly split from the meson sector of the spectrum in the chiral limit.

\begin{figure}
\begin{center}
\scalebox{0.55}{\includegraphics[angle=0,width=1.0\textwidth]{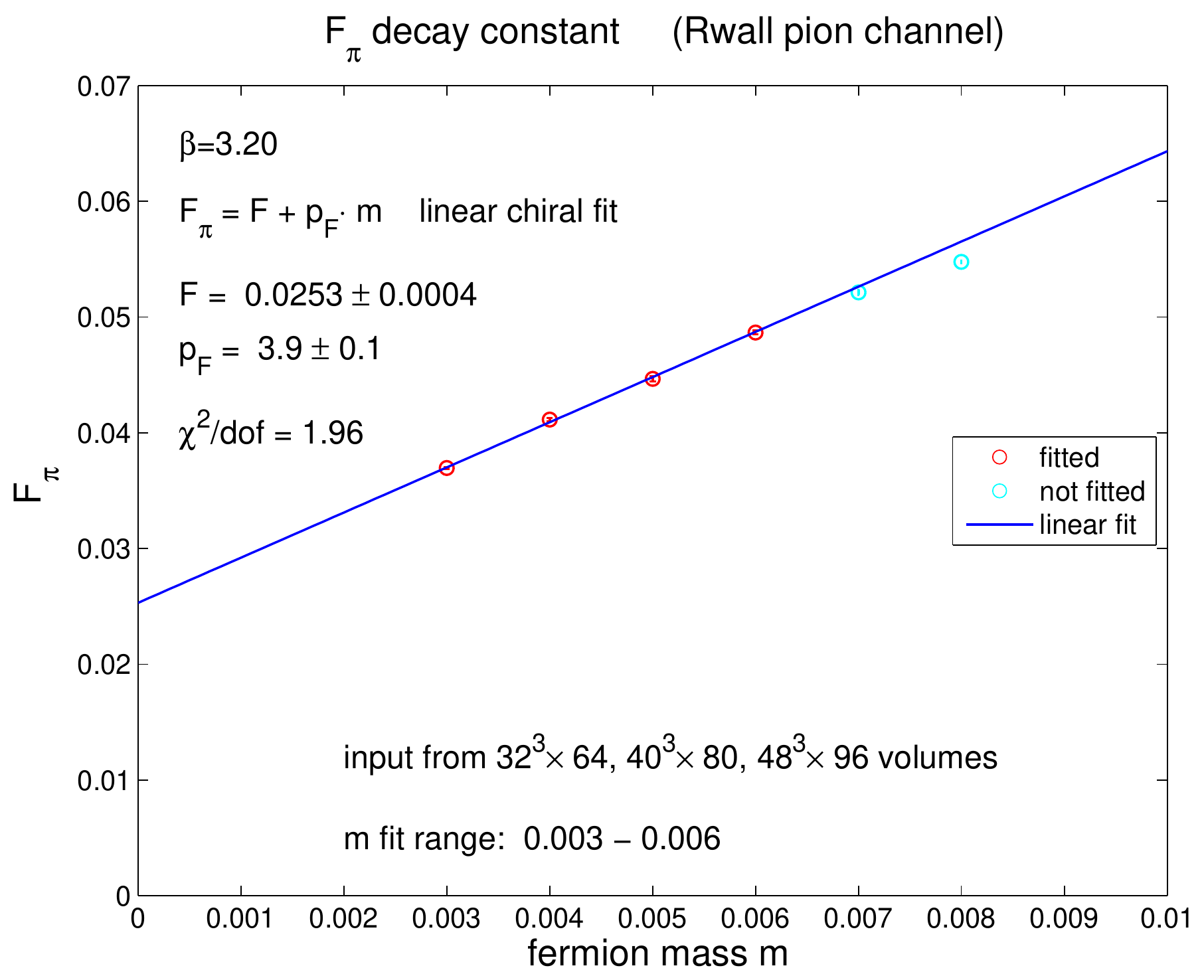}}
~ ~ ~
\scalebox{0.18}{\includegraphics[angle=0,width=1.0\textwidth]{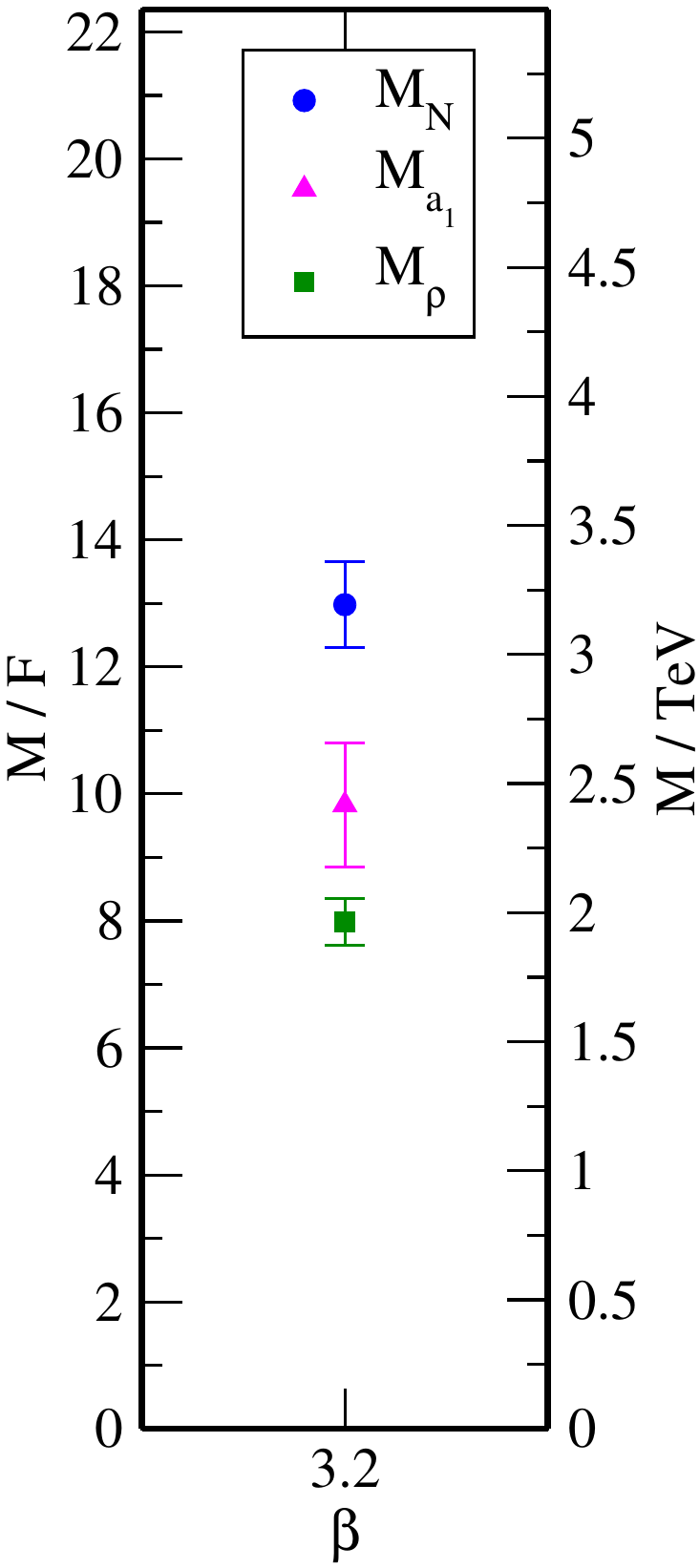}}
\vspace{-0.3cm}
\caption{{\footnotesize Chiral extrapolation of $F_\pi$. The simulations are performed on lattices with
$\beta=3.20$, $V=48^3 \times 96$ for at fermion mass $m=0.003$ and V$=32^3 \times 64$ for the heavier fermion masses, 
using 200 to 300 configurations, see \cite{ourfuture} for more details (left). Hadron spectroscopy at $\beta=3.20$ converted to physical units (right).}}\label{physical}
\end{center}
\end{figure}

To convert to physical units, one can use the scale set by the chiral limit of the Goldstone boson decay constant $F_\pi$, denoted by $F$.  The left plot of figure \ref{physical} shows the chiral
extrapolation of $F_\pi$, measured on lattices of size $48^3 \times 96$ for $m=0.003$ and $32^3 \times 64$ for the heavier fermion masses. A detailed description of the analysis is given in~\cite{ourfuture}, the result we use here is $F=0.0253(4)$. The value in the chiral limit is likely to evolve, due to additional lattice simulations pushing to even lighter fermion mass, as well as modifications of the analysis taking into account lattice artifacts. The BSM implementation of the theory identifies the chiral limit of the Goldstone boson decay constant with the value of the scalar vacuum expectation value in the Standard Model, namely $F= v_R = 246$~GeV. This requirement follows from the generation of the Electroweak gauge boson mass $m_W = (g F)/2$ due to Goldstone boson contributions to the gauge boson vacuum polarization.
With this conversion, the baryon in the chiral limit is at approximately $3$~TeV. The nucleon, vector and axial vector meson masses are shown in physical units on the right panel of figure \ref{physical}. Our initial studies indicate a composite scalar in the sextet model which is light, with $M_{\rm H}/F$ in the range 1 to 3 in the chiral limit~\cite{Fodor:2015vwa}. The large uncertainty is due to the difficulty of extracting a state with vacuum quantum numbers, which requires disconnected fermion diagrams to be calculated, a notoriously challenging computational problem. The lightness of the scalar, far separated from the remainder of the spectrum, means in the BSM context that experimental studies would need to explore the few TeV range for this model to be critically tested~\cite{Aad:2015yza}.  

It has recently been reported by the ATLAS and CMS collaborations that there is an observed excess in $WW, WZ$ and $ZZ$ diboson pairs at around 2 TeV~\cite{Aad:2015owa,CMS:2015gla}. One possible explanation is a vector resonance, which in our context would be the $\rho$ state. As indicated in figure \ref{physical}, the simulation results predict that the vector state is at roughly 2 TeV in the sextet model. This prediction will be refined in our ongoing study, including the dependence on the lattice cutoff, and will become of growing importance if the experimental excess persists. In Section 3 we already commented briefly on the recently found 
diphoton excess in ATLAS and CMS resonance searches~\cite{diphoton:2015}.

\vspace{-0.3cm}

\section{Summary and outlook}

\vspace{-0.2cm}

We have developed the operator technology to extract baryon states using staggered lattice fermions in the sextet model. The first results are encouraging that the lightest baryon mass can be nailed down with good precision from the current generation of lattice simulations. The emerging picture is of a spectrum with the baryon significantly above the vector and axial vector mesons in the chiral limit. The next step will be to extend the analysis, allowing more control over the chiral extrapolation and removing the distortion of the spectrum due to lattice artifacts and possible finite-volume contamination.

The study of the baryon is a natural extension of our ongoing work to explore the meson spectrum of the sextet model and shore up the case that the theory is near-conformal, with a massive spectrum in the chiral limit. Whether or not the model is ultimately viable will largely hinge on the fate of the composite scalar, which should be light for the minimal model to dynamically generate a Higgs impostor. The embedding of the sextet theory into the Standard Model brings other facets into play, such as constraints on the relic abundance of sextet model baryons in the early Universe, and experimental limits on stable particles with fractional charge~\cite{Perl:2009zz}. The scenario of a neutral sextet baryon with additional lepton doublets for anomaly cancellation shows the possibility for rich BSM physics beyond simply a composite Higgs state, and will be brought into sharp relief if the model continues to warrant future study.  

\vspace{-0.3cm}

\section{Acknowledgement}

\vspace{-0.2cm}


We acknowledge support by the DOE under grant DE-SC0009919,
by the NSF under grants 0970137 and 1318220, by the DOE ALCC award for the BG/Q Mira platform
of Argonne National Laboratory, by OTKA under the grant OTKA-NF-104034, and by the Deutsche
Forschungsgemeinschaft grant SFB-TR 55. Computational resources were provided by the Argonne Leadership Computing Facility under an ALCC award, by USQCD at Fermilab, 
by the NSF XSEDE program, by the University of Wuppertal, by Juelich Supercomputing Center on Juqueen
and by the Institute for Theoretical Physics, Eotvos University. 
We are grateful to Szabolcs Borsanyi for his code development for the BG/Q platform. We are also 
grateful to Sandor Katz and Kalman Szabo for their code developent for the CUDA platform \cite{Egri:2006zm}. KH wishes to thank the Institute for Theoretical Physics and the Albert Einstein Center for Fundamental Physics at the University of Bern for their support.


\end{document}